\documentclass[%
 aip,
 pop
 amsmath,amssymb,
 reprint,%
]{revtex4-1}

\usepackage{graphicx}
\usepackage{dcolumn}
\usepackage{bm}

\newcommand{\sss}{\scriptscriptstyle}
\newcommand {\be}{\begin{equation}} 
\newcommand{\ee}{\end{equation}}    

\def\vti{v_{{\sss T}i}}
\def\ve{v_{{\sss T}e}}
\def\vte2{v_{{\sss T}\alpha}}

\def\vtj{v_{{\sss T}j}}

\def\vtse{v_{{\sss T}se}}
\def\vtsi{v_{{\sss T}si}}

\def\vtfe{v_{{\sss T}fe}}
\def\vtfi{v_{{\sss T}fi}}

\begin{document}
\preprint{AIP/123-QED}

\title[]{Ion plasma wave and its instability in interpenetrating  plasmas}

\author{J. Vranjes}
  \email{jvranjes@yahoo.com}
  \affiliation{
Institute of Physics Belgrade, Pregrevica 118, 11080 Zemun, Serbia\\
}%
\author{M. Kono}%
 \email{kono@fps.chuo-u.ac.jp}
\affiliation{
Faculty of Policy Studies, Chuo University, Tokyo, Japan
}%


\date{\today}

\begin{abstract}
Some essential features of the ion plasma wave in both kinetic and fluid descriptions are presented. The wave develops at
wavelengths shorter than the electron Debye radius. Thermal  motion of electrons at this scale is such that they overshoot the electrostatic potential perturbation  caused by ion bunching, which consequently propagates as an unshielded wave, completely unaffected by electron dynamics. So in the simplest fluid description, the electrons can be
taken as a fixed background. However, in the presence of magnetic field and for the  electron gyro-radius shorter than the Debye radius, electrons can participate in the wave and can increase its damping rate. This is determined by the ratio of the electron gyro-radius and the Debye radius.  In interpenetrating plasmas (when one plasma drifts through another), the ion plasma wave can easily become growing and this growth rate is quantitatively presented for the case of an argon plasma.
\end{abstract}

\maketitle



\section{\label{s1} Introduction }
%

 The ion plasma (IP) mode  has been predicted long ago \citep{ses, kur} and experimentally confirmed in \citet{ba,ba2} (but   see also in
   \citet{ses1}, \citet{ver}, \citet{dg}, \citet{kra}, \citet{arm}). Yet, the mode  is not frequently  discussed  in textbooks (as  exceptions see  \citet{baum} and \citet{bit}), probably mostly because it is expected  to be strongly damped  \citep{be} and therefore of no practical importance. In fact, in most of the books the curve depicting its  low frequency counterpart, i.e., the longitudinal ion acoustic (IA) wave,  saturates towards the ion plasma frequency $\omega_{pi}$, and the longitudinal mode is absent above it.  Such a profile of the IA wave is simply the result of the standardly assumed quasi-neutrality in the perturbed state.   However,   the quasi-neutrality condition is not necessarily satisfied for short wavelengths, and the  line depicting the longitudinal mode indeed continues even above $\omega_{pi}$. The mode in this frequency regime may play an important role in plasmas with multiply ionized ions and where in the same time  $z_i T_e/T_i\gg 1$, when its damping is considerably reduced; here $z_i$ is the ion charge number.
 Particle-in-cell simulations of this space-charge wave at the ion plasma frequency  are performed by  \citet{jon} in order to accelerate particles to high energies.
 More recently, the IP mode has been discussed by \citet{dra}, where an intense pump wave is used to  increase the frequency of an ordinary ion-acoustic wave to frequencies
near the ion-plasma frequency, and the wave obtained in such a manner is called an  induced IP  wave.

The physics of the ion plasma  mode is very briefly described by \citet{be}, and \citet{dra}. For wavelengths satisfying the condition $k \lambda_{de} \geq 1$, where $\lambda_{de} $ is the electron Debye radius, the electron chaotic motion is such that they cannot effectively shield charge fluctuations at so short wavelengths. The electrons simply
 overshoot potential perturbations caused by ion  oscillations and they consequently act only as a background (which is neutralizing only in the equilibrium, but not also in the perturbed state).  The electron and ion role is thus reversed as compared  to the electron (Langmuir) plasma wave.  Compared with the ion acoustic  wave, where the perturbed thermal pressure acts as the restoring force, in the  ion plasma wave  the  Debye shielding is ineffective and the restoring force is the electrostatic Coulomb force. So  {\em the ion density} oscillates at {\em the ion-plasma frequency} instead of at the  acoustic frequency.

  The absence of the Debye shielding associated with the IP wave may affect various other phenomena in plasmas \citep{dra}, like Stark broadening, braking radiation processes, atomic transitions, etc.  Therefore identifying possible sources of the IP mode (like in the case of interpenetrating plasmas, see later in the text), is of great importance and it can help in correctly interpreting observations in  space and astrophysical plasmas.

  On the other hand, taking the solar wind parameters \citep{baum} at 1 AU ($T_e=1.5\cdot 10^5$ K, $T_i=1.2\cdot 10^5$ K, $n_0\approx 5\cdot 10^6$ m$^{-3}$) reveals that the Debye radius for both electrons and protons is around 11 m. The wavelengths of the IP wave are below $2 \pi \lambda_{de}$ and thus eventually exceeding the typical size of   space satellites, that are of the size of a fridge or bigger (c.f., Messenger Mercury probe is about $1.3  \times 1.4  \times 1.8$ m, while Ulysses solar probe is around $3.3 \times 3.2 \times 2.1$ m size). It is thus obvious that in such an environment the satellite may appear immersed in the unshielded  electric field of a passing IP wave. A growing IP wave can consequently affect the performance of  its electric installations, and can lead to surface charging and sparks.

It is important to stress that both  modes (IA and IP) belong to the same longitudinal branch of a dispersion equation, yet with different wavelengths $k\lambda_{de}<1$ and     $k\lambda_{de}>1$, respectively.
 The physics of the longitudinal electrostatic wave in these two wavelength limits is completely different, as described above, which   justifies the two different names (IA and IP) being used.

\section{\label{sec2} Basic properties  of the ion plasma wave}
\subsection{Kinetic description}
 In the usual  electron-proton plasma and in the limit
 \be
k \vti\ll \omega\ll k \ve,\label{e}
\ee
 where $\vti, \ve$ denote the ion and electron thermal speeds $(\kappa T_j/m_j)^{1/2}$, the kinetic dispersion equation which yields both the ion plasma wave and the ion acoustic  wave  reads \citep{be, v1}:
\[
\Delta\equiv 1+ \frac{1}{\lambda_{de}^2 k^2} - \frac{\omega_{pi}^2}{\omega^2} - \frac{3 \omega_{pi}^2 k^2 \vti^2}{\omega^4}
\]
\be
+ i
(\pi/2)^{1/2} \left\{\frac{\omega_{pe}^2 \omega}{k^3
\ve^3} + \frac{\omega_{pi}^2 \omega}{k^3 \vti^3} \exp\left[-
\omega^2/(2 k^2 \vti^2)\right]\right\}=0. \label{e1} \ee
Here, $\omega_{pj}^2=e^2 n_0/(\varepsilon_0 m_j)$ are the electron and ion plasma frequencies in an equilibrium quasi-neutral plasma with the equilibrium number density $n_0$, and $\lambda_{dj}=\vtj/\omega_{pj}$ is the Debye radius of the species $j$. The IP wave is obtained for the small second term in Eq.~(\ref{e1}), and the frequency and the damping  rate are approximately given by
\be
\omega_r^2=\omega_{pi}^2\left(1+ 3 k^2\lambda_{di}^2\right), \label{e2}
\ee
\[
  \gamma=-\left(\frac{\pi m_e}{8m_i}\right)^{1/2}\!\!\frac{\omega_{pi}}{ k^3\lambda_{de}^3}\left(1+
     3 k^2 \lambda_{di}^2\right)^2
     \]
     \be
     \times
     \left[1
         + \left(\frac{T_e}{T_i}\right)^{3/2}\!\!\left(\frac{m_i}{m_e}\right)^{1/2}\!\!
  \exp\left(-\frac{3}{2}-\frac{1}{2k^2\lambda_{di}^2}\right)\right].
\label{e3}
\ee
  For plasma with hot protons $T_i=T_e$ and for $ k^2\lambda_{de}^2>1$ this yields
  \[
  \gamma \simeq -\left(\frac{81\pi }{8}\right)^{1/2} k\vti\left[\left(\frac{m_e}{m_i}\right)^{1/2}\!\!\! + \exp\left(-\frac{3}{2}\right)\right],
  \]
  so  that $|\gamma/\omega_r|\simeq 0.8$. The damping of the IP mode is mainly due to the ion contribution (the second term in the bracket), and it is  stronger than the damping of  the ion acoustic wave in plasmas with hot ions. It should be stressed here that the dispersion equation (\ref{e2}) can also be derived  if electrons are taken as a completely fixed background. This will be discussed in more detail in the fluid description later in the text.

\begin{figure}
 \centering
\includegraphics[height=6cm,bb=18 15 274 216,clip=]{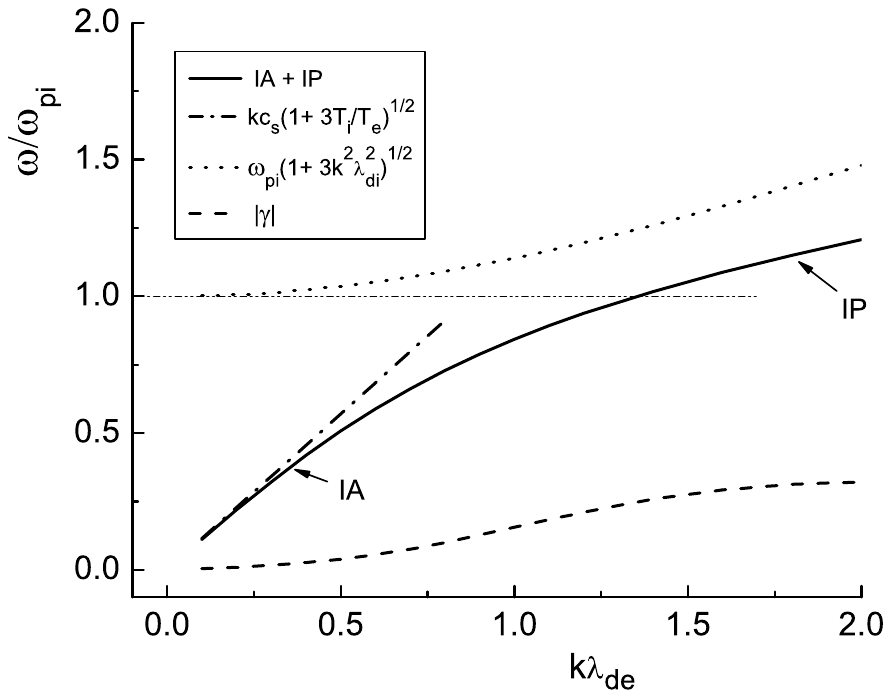}
\caption{\label{f1}Transition from the ion acoustic (IA) wave for small $k\lambda_{de}$,  to the ion plasma (IP) wave  for $k\lambda_{de}\geq 1$ in an argon plasma, with both modes being Landau-damped.   }
\end{figure}

In Fig.~\ref{f1} we give the spectrum from the general equation (\ref{e1}) for an arbitrary argon plasma with the following parameters: $n_0= 10^{15}$ m$^{-3}$, $T_e= 5\cdot 10^4$ K, $T_i=0.1\, T_e$.
The real part of the frequency (full line in the figure) is obtained  from the real part of Eq.~(\ref{e1}) and it reads
\be
\omega_r^2=\frac{k^2 c_s^2}{2 (1+ k^2 \lambda_{de}^2)} \left\{1\pm \left[1+ \frac{12 T_i}{T_e}\left(1+ k^2 \lambda_{de}^2\right)\right]^{1/2}\right\}.\label{f}
\ee
Here, $c_s^2=\kappa T_e/m_i$, and only the sign plus should be kept for physical reasons.
The damping (dashed line in the figure) follows from $\gamma\simeq -\mbox{Im} \Delta/(\partial \mbox{Re} \Delta/\partial \omega)_{\omega\equiv \omega_r}$, which becomes
\[
\gamma=-\left(\frac{\pi}{8}\right)\frac{\omega_r^4}{k^3\ve^3}\left(\frac{m_i}{m_e}\right)^{3/2}\!\!\!\!\frac{1}{1+ 6 k^2 \vti^2/\omega_r^2}
\]
\be
\times
\left[\left(\frac{m_e}{m_i}\right)^{1/2} +
\left(\frac{T_e}{T_i}\right)^{3/2}\exp\left(-\frac{\omega_r^2}{2 k^2 \vti^2}\right)\right].
\label{d}
\ee
The figure shows the presence of the wave (real $\omega_r$) even above the ion plasma frequency $\omega_{pi}$.  The corresponding (normalized) Landau damping (\ref{d}), the  dashed line,  in the  IA range  of the dispersion curve is very small as compared with the damping of the IP part. For example $\gamma=-0.005$ at $k\lambda_{de}=0.1$ (which corresponds to the IA part of the spectrum) while $\gamma=-0.26$ at $k\lambda_{de}=1.4$ (corresponding to the IP wave part).  Note that $|\gamma|/\omega_r=0.04$, and $|\gamma|/\omega_r=0.26$ for the two cases, respectively, so the relative damping of the IP wave is greater by about a factor 6.5. For greater ion temperature the damping should be much greater, however this cannot be properly checked within the present approximate derivations  because the conditions $|\gamma|\ll \omega_r$ and $k \vti\ll \omega_r$, used in the derivation of the  dispersion equation (\ref{e1}) and the damping (\ref{d}),  become strongly violated, so that the general dispersion equation [see Eq.~(\ref{e4}) below] must be solved exactly instead of using equation (\ref{e1}). But it is obvious that in normal circumstances (e.g., for a simple electron-ion plasma)  the IP mode is indeed unlikely to appear without an efficient source.

\begin{figure}
 \centering
\includegraphics[height=6cm,bb=18 16 275 214,clip=]{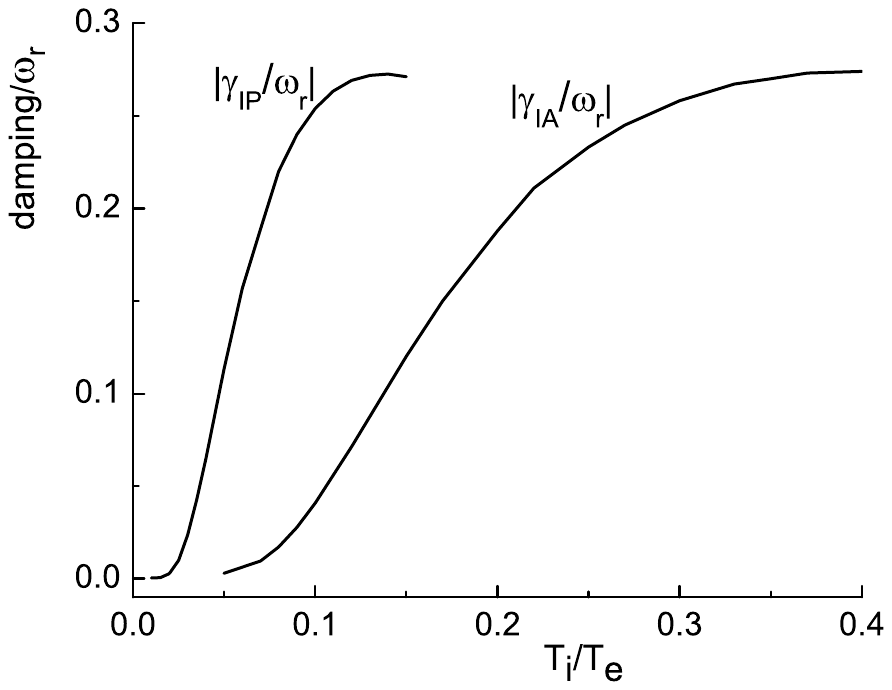}
\caption{\label{f2}Absolute value of the  ion acoustic  and the ion plasma mode damping in terms of ion-electron temperature ratio.   }
\end{figure}

In Fig.~\ref{f2} we give the damping of the IA mode at $k\lambda_{de}=0.1$, and of the IP mode at $k\lambda_{de}=1.4$ (see Fig.~\ref{f1}) in terms of the temperature
ratio $T_i/T_e$ for the same density as in Fig.~\ref{f1}.  The apparent saturation (and decrease) of the damping for the IA wave for  $T_i/T_e>0.4$,
and for $T_i/T_e>0.15$ in the case of the IP wave, is only due to the fact that the condition (\ref{e}) becomes violated for greater ion temperature.  So above the given temperature ratios  the expansion of the plasma dispersion function is not justified, and in the same time the term $6 k^2 \vti^2/\omega_r^2$ in the denominator of Eq.~(\ref{d}) becomes greater than 1 and affects the line shape. The saturation seen in Fig.~\ref{f2} is thus not physical; in reality, the damping of both waves would further increase, and the  IP wave damping is expected to be stronger than the IA wave damping. This can only be demonstrated by solving the plasma dispersion function numerically for the large $T_i/T_e$ ratio. For the IA wave such a saturation and its comparison with the exact damping may be seen in \citet{ch, vld1}, and \citet{vld2}.

\subsection{Effects of collisions with neutrals within the kinetic theory}

Derivations can easily be generalized to include collisions with neutrals. We may start from the kinetic equation  with the model collisional integral
\be
\frac{\partial f_j}{\partial t} +   \vec v\!\cdot\! \frac{ \partial f_j}{\partial \vec r}
  + \frac{q_j \vec E}{m_j} \!\cdot\! \frac{\partial f_j}{\partial
\vec v}
=-\nu_{jn} \left(f_j- n_j f_{jn}\right). \label{c1}
\ee
Here
\be
f_{jn}\!=\!\left(\!\frac{m_j}{2\pi \kappa T_{jn}}\!\right)^{3/2}\! \!\!\!\!\exp\!\left(\!-\frac{m_jv^2}{2 \kappa T_{jn}}\!\right), \!\quad T_{jn}\!=\!\frac{m_j T_n + m_n T_j}{m_j + m_n}.\label{c2}
\ee
Stationary equilibrium is satisfied by $f_{j0}=n_{j0} f_{jn}$.  In the perturbed state we have
\be
\frac{\partial f_{j1}}{\partial t} +   \vec v\!\cdot\! \frac{ \partial f_{j1}}{\partial \vec r}
  + \frac{q_j \vec E_1}{m_j} \!\cdot\! \frac{\partial f_{j0}}{\partial
\vec v}
=-\nu_{jn} \left(f_{j1}- n_{j1} f_{jn}\right). \label{c3}
\ee
 To simplify notation we shall assume that the ion and neutral mass and temperature  are equal, and for electrons after neglecting the terms with the mass ratio we have $T_{en}=T_e$. The perturbed
 number density becomes
\be
n_{j1}=-\frac{q_j n_{j0} \phi_1}{\kappa T_j} \frac{\left[1- Z\left(\frac{\omega_1}{k\vtj}\right)\right]  }{1- \frac{i \nu_{jn}}{\omega_1}Z\left(\frac{\omega_1}{k\vtj}\right)  }, \quad \omega_1=\omega+ i \nu_{jn}. \label{c3}
\ee
This may be used in the Poisson equation to obtain the dispersion equation. In the frequency range $\vti\ll |\omega_1/k|\ll \ve$ we use  the same expansions as earlier.  The $Z(b_e)$ function in the denominator on the right hand side yields an imaginary term which in the combination with the rest of the term in the denominator yields a negligible real correction. For ions, assuming that $|\omega|>\nu_{in}$, from the $Z_{bi}$ function in the denominator it is enough to keep the first term in the expansion. The damping for the electron-ion case with dominant  ion-neutral collisions consequently becomes:
\[
\gamma=-\left(\frac{\pi}{8}\right)^{1/2}\!\! \frac{\omega_r^4}{k^3\ve^3}\left(\frac{m_i}{m_e}\right)^{3/2}\!\!\!\!\frac{1}{1+ 6 k^2\vti^2/\omega_r^2}\left[\left(\frac{m_e}{m_i}\right)^{\!1/2}\!\!+  \right.
\]
\be
\left.
\left(\frac{2 }{\pi}\right)^{\!1/2}\!\!\frac{k^3c_s^3}{\omega_r^3} \frac{\nu_{in}}{\omega_r}+\left(\frac{T_e}{T_i}\right)^{3/2} \!\! \!\!\exp\left(-\frac{\omega_r^2}{2 k^2 \vti^2}\right)\right].
\label{c4}
\ee
The real part of frequency is the same as before. The damping given by Eq.~(\ref{c4}) applies to both the IP and IA wave. From the second term in the bracket in Eq.~(\ref{c4}), it may be concluded that the collisions will affect the IP and IA modes differently. This is because the ratio $\nu_{in}/\omega_r$ is normally smaller for the IP wave, though  the ratio $k^3c_s^3/\omega_r^3\equiv (c_s/v_{ph})^3$, $v_{ph}=\omega_r/k$,  may be greater for the IP mode. As example, for parameters from Fig.~\ref{f2} we have $c_s/v_{ph}=1.37$ for the IP and $c_s/v_{ph}=0.9$ for the IA wave, while at the same time $\omega_{{\sss IP}}\simeq 9 \omega_{{\sss IA}}$.

\subsection{Ion plasma wave within  fluid theory}

Derivation within the fluid theory is  straightforward and this will be presented  here only in order to highlight some essential differences
between the IP wave and IA wave regarding the effects of electron-ion collisions, and the general role of electrons. We start from the momentum equations with the BGK collisional terms [which can directly be derived from the kinetic equation similar to (\ref{c1}) assuming e-i collisions  only]:
\[
m_i n_i \left(\frac{\partial}{\partial t} + \vec v_i\cdot \nabla\right)\vec v_i= - e n_i \nabla \phi- \kappa T_i \nabla
n_i
\]
\be
+ m_e n_e \nu_{ei} (\vec v_e - \vec v_i),\label{s1} \ee
\be
0=  e n_e \nabla \phi- \kappa T_e \nabla
n_e - m_e n_e \nu_{ei} (\vec v_e - \vec v_i),\label{s2} \ee
and the continuity equation
\be
\frac{\partial n_j}{\partial t}+ \nabla\cdot(n_j \vec v_j)=0, \quad {j=e, i}. \label{s4} \ee
One interesting feature related to the IA wave and collisions is immediately seen. From linearized Eq.~(\ref{s4}) we have $v_{i1}=\omega n_{i1}/(k n_0)$, $v_{e1}=\omega n_{e1}/(k n_0)$. Using this in Eqs.~(\ref{s1}, \ref{s2}) it is obvious
that the friction force terms exactly vanish if the quasi-neutrality condition is used in the perturbed state. This approximation is appropriate for the IA wave, hence  this mode will be unaffected by electron-ion collisions  \citep{v3, v4}. If the Poisson equation is used instead of quasi-neutrality, the IA wave damping formally appears \citep{v3, v4}, $\gamma_{\scriptscriptstyle{IA}}=i \nu_{ei} (m_e/m_i) k^2 \lambda_{de}^2$, but it is clearly negligible because  $k^2 \lambda_{de}^2\ll 1$.

On the other hand, for the IP wave the  Poisson equation must be used. For the perturbed electron number density we obtain
\be
 n_{e1}=\frac{e n_0}{\kappa T_e}\left(1-\frac{i\nu_{ei} \omega}{\omega_{pe}^2}\right)\phi_1. \label{s5}
 \ee
The ion number density becomes:
\be
n_{i1}=\frac{e n_0k^2}{m_i\omega^2}\,\frac{1-i\nu_{ei} \omega/\omega_{pe}^2}{1-k^2 \vti^2/\omega^2}\,  \phi_1.\label{s6}
\ee
These two expressions are used in the Poisson equation $\varepsilon_0\nabla^2\phi_1=-e(n_{i1}- n_{e1})$ which, on conditions $k^2\lambda_{de}^2>1$, $T_e>T_i$ yields the IP wave spectrum
\be
\omega\simeq \pm \omega_{pi}\left(1+ k^2 \lambda_{di}^2\right)^{1/2} - \frac{i\nu_{ei}}{2} \frac{m_e}{m_i}.\label{s7}
\ee
This damping is clearly much stronger than the corresponding damping of the IA wave given above. However, for the parameters used in Figs.~\ref{f1},~\ref{f2}, this collisional damping is completely negligible as compared with the kinetic damping of the IP wave presented in the figures.

The shape of Eq.~(\ref{s7}) is formally similar to the electron plasma wave, which has a cut-off (reflection) at the electron plasma frequency. However,   Eq.~(\ref{s7})
is just an approximate expression and the cut-off is not an actual feature of the wave; the dispersion curve of the mode is a smooth line which crosses $\omega_{pi}$ as depicted in Fig.~\ref{f1}, and going into the limit $k\rightarrow 0$ and  $\omega \rightarrow\omega_{pi}$ is not physically justified.

Interestingly, electron dynamics does not need to be included in the fluid equations to obtain the basic properties of IPW described above. Observe first that the full electron response described by
 Eq.~(\ref{s5}) is the result of the necessity  to formally satisfy momentum conservation due to friction between the two species.  Second, the damping term in Eq.~(\ref{s7}) can be written in an equivalent form $-i \nu_{ie}/2$.  Third, such a dispersion equation with $\nu_{ie}$ can easily be obtained if the electron dynamics is completely neglected in the first place (i.e., if they are treated as a fixed background),  and if the friction term in the ion momentum equation (\ref{s1}) is taken in the shape $-m_i n_i \nu_{ie}\vec v_i$. In other words, the electron dynamics plays no role in the physics of the IP wave, and   in the limit  $k^2\lambda_{de}^2>1$ used here all electron terms vanish in any case.
So a self-sufficient set of equations  needed to describe the basic properties of the IP mode in the fluid theory  reads:
\be
m_i n_i \left(\frac{\partial}{\partial t} + v_i \frac{\partial}{\partial x}\right)v_i= - e n_i \frac{\partial \phi}{\partial x}- \kappa T_i\frac{\partial n_i}{\partial x} -m_i n_i \nu_{ie}v_i,
\label{ne}
\ee
\be
  \frac{\partial n_i}{\partial t}+ \frac{\partial}{\partial x}(n_i v_i)=0, \quad \varepsilon_0 \frac{\partial^2 \phi}{\partial x^2}=-e n_i.\label{ne1}
  \ee
These ion equations can further be generalized  by inclusion of various additional phenomena yet these will not include the dynamics of electrons in any case.   In the nonlinear theory, the   reductive perturbation method can be used for the nonlinear Eqs.~(\ref{ne}, \ref{ne1}) for the IP mode  in the manner similar to  the IA wave soliton theory \citep{sm,sw}.

\subsubsection{\label{sec5} Effects of  magnetic field}

 Within the frequency range of the IP wave, the shortest spatial scale for electron response is defined by the electron  Debye radius, and due to this they play no role in the sub-Debye IP wave dynamics. However, in the presence
 of an external magnetic field, there appears the electron gyro-radius $\rho_e=\ve/\Omega_e$, $\Omega_e=e B_0/m_e$, as yet another characteristic length parameter.   If $\rho_e>\lambda_{de}$  the electrons will be unmagnetized  for the IP wavelengths of interest, and the electron dynamics will remain unimportant as if the magnetic field is absent. In this case, obviously the ions will be  unmagnetized as well, and the magnetic field will play no role. The IP wave will propagate  at any angle with respect to the magnetic field, it will be unaffected by its presence, and the wave will still completely be determined by  ion dynamics only.
 But if $\rho_e<\lambda_{de}$, the electrons may  become magnetized within the wavelength of the IP wave, and they will consequently be able to respond to the IP wave electric field by performing the usual  $\vec E\times \vec B$-drift.

 Taking the magnetic field $B_0=0.1$ T and other parameters as in Fig.~1, reveals that  the ion-gyro radius is  about 9 times larger than $\lambda_{de}$, while at the same time the electron gyro-radius
is ten times shorter than the electron Debye radius. Hence, ion dynamics in the IP wave will remain unchanged, but electrons are magnetized and they might modify the IP wave (this will be checked  further in this section).  In the case of solar corona (e.g., with plasma density $5\cdot 10^{14}$ m$^{-3}$ and temperature $10^6$ K), taking the magnetic field $B_0=10^{-4}$ T, and $B_0=10^{-3}$ T, yields $\rho_i/\lambda_{de}\simeq 3000$ and $\rho_i/\lambda_{de}\simeq 300$, respectively.  Similarly, for electrons $\rho_e/\lambda_{de}\simeq 71$ and $\rho_e/\lambda_{de}\simeq 7$, respectively. Consequently, for such magnetic fields in the corona  both ions and electrons  will remain unmagnetized for the IP wavelengths, and the wave will remain unaffected by the magnetic field and by electrons, regardless of the direction of propagation.

To see this all in detail, we may take the initial magnetic field as $\vec B_0=B_0\vec e_z$ and assume the IP wave propagating in any direction, determined by the wave vector $\vec k$.
From what we have seen so far it is clear that, if $\vec k$ has one component along $\vec B_0$, the electron dynamics along   $\vec B_0$ will play no role (the assumed wave-lengths are below the electron response scale). Hence, to see how the magnetic filed introduces the electron dynamics into the IP wave, it is best to study  the IP wave propagating  strictly perpendicular to $\vec B_0$, say in the $x$-direction, i.e,  $\vec k=k\vec e_x$. Without collisions, in such a geometry the ion motion is in $x$-direction, while electrons perform  drift motion in $y$-direction. Collisions couple the two motions, as will be shown below.

The required electron momentum equation now reads:
\be
0=  e n_0 \nabla \phi- \kappa T_e \nabla
n_{e1} - e n_0 \vec v_{e1}\times \vec B_0 - m_e n_0 \nu_{ei} (\vec v_{e1} - \vec v_{i1}),\label{s2a} \ee
while the continuity equation for the two species reduces to
\be
\frac{\partial n_{j1}}{\partial t}+ n_0 \frac{\partial v_{jx1}}{\partial x}=0. \label{s4a} \ee
Instead of Eq.~(\ref{s5}) we now have
\be
 n_{e1}=\frac{e n_0}{\kappa T_e}\frac{1-i \nu_{ei}\omega/\omega_{pe}^2}{1+ m_e/(m_i k^2\rho_e^2) - i\omega/(\nu_{ei} k^2\rho_e^2) }\phi_1. \label{s5a}
 \ee
For the ions, the $x$-component of the momentum is the same as their dynamics from the previous section so that the ion number density is again given by Eq.~(\ref{s6}), although they have an additional motion in the $y$-direction caused by the electron drag, i.e.,  $-i m_i \omega v_{iy}=m_e \nu_{ei} (v_{ey}-v_{iy})$. This ion component affects the electron dynamics, and it is already used in the derivation of the electron number density (\ref{s5a}).
Eqs.~(\ref{s6}, \ref{s5a}) are used in the Poisson equation which yields the dispersion equation
\[
1=\frac{\omega_{pi}^2}{\omega^2} \frac{1- i \nu_{ei} \omega/\omega_{pe}^2}{1- k^2 \vti^2/\omega^2}
\]
\be
- \frac{1}{k^2\lambda_{de}^2} \frac{1- i \nu_{ei} \omega/\omega_{pe}^2} {1+ m_e/(m_i k^2\rho_e^2) - i \omega/(\nu_{ei} k^2\rho_e^2)}. \label{cc1}
\ee
The contribution of $m_e/(m_i k^2\rho_e^2)$ in the second (electron) term is clearly negligible. The rest of this equation can be
written as
\[
1\!=\!\frac{\omega_{pi}^2}{\omega^2} \frac{1- i \nu_{ei} \omega/\omega_{pe}^2}{1- k^2 \vti^2/\omega^2} - \frac{1}{k^2\lambda_{de}^2} \frac{1\!+ \!a - (i \nu_{ei} \omega/\omega_{pe}^2) (1- c)} {1+ b},
\]
\[
a\equiv \frac{\omega^2}{\omega_{pe}^2}\frac{1}{k^2\rho_e^2}, \quad b\equiv \frac{\omega^2}{\nu_{ei}^2}\frac{1}{k^4\rho_e^4}, \quad  c\equiv \frac{\omega_{pe}^2}{\nu_{ei}^2}\frac{1}{k^2\rho_e^2},
\]
and it can be discussed in various ways.
As an example, we may assume $a>1$, $b>1$, $c>1$.
All three of these conditions contain the assumed fact that electrons are magnetized. The dispersion equation now becomes
\be
(\omega^2 - k^2 \vti^2) (1+ d) \!=\!\omega_{pi}^2 - i \nu_{ei} \omega \!\left[\!\frac{m_e}{m_i} + \frac{\rho_e^2}{\lambda_{de}^2}\!\left(\!1- \frac{k^2\vti^2}{\omega^2}\!\right)\!\right],
\label{s8}
\ee
\[
d= \frac{\rho_e^2}{\lambda_{de}^2}\frac{\nu_{ei}^2}{\omega_{pe}^2}.
\]
Here, $d$ can also be omitted, and in the imaginary term $k^2\vti^2/\omega^2<1$, and in addition typically $m_e/m_i< \rho_e^2/\lambda_{de}^2$. Thus  the imaginary term
finally reduces to  $-i \nu_{ei} \omega \rho_e^2/\lambda_{de}^2$, while the IP wave frequency remains practically the same as before. Comparing this imaginary part with Eq.~(\ref{s7}) it is seen that, under assumptions used above, the damping in the presence of magnetic field is considerably increased. This is due to the fact that in the present case the ions are subject to additional movement (perpendicular to the wave propagation direction) caused by the  electron drag in the direction of the $\vec E\times \vec B$-drift.

\section{\label{sec3} Instability of ion plasma wave}

  The IP wave  can become unstable  in case of interpenetrating/permeating plasmas, that is when one quasi-neutral plasma propagates through another quasi-neutral static plasma. For the ion acoustic wave this has been shown in \citet{v1}, and in \citet{v2}. Examples of such plasmas in space and astrophysical environment are numerous, to mention just a few.
    a) Observations reported in \citet{be} and \citet{bru} reveal jets generated in the transition region of the solar atmosphere with the rate of about 24 events per second throughout the solar atmosphere, and having a few thousand kilometers in diameter. The jets were  moving upwards with speeds of around 400 km/s.
    b) Similarly, upflows of plasma in numerous chromospheric spicules   may reach any height in the corona between $5\cdot 10^3$ and $2\cdot 10^5$ km. Their  diameters are typically greater than 1000 km, and they cover a few percent of the solar atmosphere  in any moment.
     c) An opposite phenomenon,  which again implies interpenetrating plasmas, is the coronal rain  \citep{be}. This is  a  plasma moving down towards and through the lower layers of the solar atmosphere with almost a free-fall speed between 50 and 100 km/s.
       d) In astrophysics, typical examples of such interpenetrating plasmas are clouds of novae and supernovae explosions propagating through the surrounding space, or solar and stellar winds moving through interstellar and interplanetary plasmas.
        e) In laboratory conditions, such interpenetrating plasmas with different parameters (e.g., pressure) can easily be realized.

         In general, there is  a free energy in such interpenetrating plasmas, which  can drive various current-less plasma instabilities; some of them related to the IA wave instability are studied in \citet{v1} and \citet{v2, vld2}.

The kinetic derivation for interpenetrating plasmas follows the procedure described in detail in \citet{v1} and it will not be repeated here. It implies Maxwellian distributions for the static plasma components (all quantities with the index $s$), and shifted Maxwellians for the electrons and ions (protons) in the plasma flowing with the speed $v_0$ (and with the corresponding quantities denoted with the index $f$). Hence, in the equilibrium we have for the two plasmas separately:
\[
n_{si0}=n_{se0}\equiv n_{s0}, \quad  n_{fi0}=n_{fe0}\equiv n_{f0}.
\]
On the other hand, in the perturbed state,  quasi-neutrality is not assumed. Instead, the derivation follows the standard procedure with the general wave equation as the starting point \citep{v1},   and for the longitudinal modes the
final dispersion equation  is
\be
1+ \sum_j \frac{1}{k^2 \lambda_j^2} [1- Z(b_j)]=0. \label{e4} \ee
Here, $b_j=(\omega- k v_{j0})/(k \vtj)$,
and
\be
Z(b_j)=\frac{b_j}{(2 \pi)^{1/2}} \int_c \frac{\exp(-
\zeta^2/2)}{b_j- \zeta} d\zeta, \label{pdf}
\ee
is the plasma dispersion function, with the integration over the
Landau contour $c$.

Using the well known expansions for $Z(b_j)$ in the case $v_{sj0}=0$, $v_{fj0}=v_{f0}$,  in the limits
\be
k \vtsi \ll |\omega|\ll k \vtse, \quad |\omega - k v_{f0}|\ll k
\vtfe, \,\, k\vtfi, \label{e5} \ee
  we obtain the dispersion equation that describes both  the IA and IP waves \citep{v1}:
\[
\Delta(k, \omega)\equiv  1+ \frac{1}{k^2 \lambda_d^2} -
\frac{\omega_{psi}^2}{\omega^2} - \frac{3 k^2 \vtsi^2
\omega_{psi}^2}{\omega^4}
\]
\[
+ i \left(\frac{\pi}{2}\right)^{1/2} \left[\frac{\omega
\omega_{pse}^2}{k^3 \vtse^3} + (\omega- k
v_{f0})\left(\frac{\omega_{pfe}^2}{k^3 \vtfe^3} +
\frac{\omega_{pfi}^2}{k^3 \vtfi^3}\right)\right.
\]
\be
\left.
 +
\frac{\omega\omega_{psi}^2 }{k^3 \vtsi^3}\, \exp\left(-
\frac{\omega^2}{2 k^2 \vtsi^2}\right)\right]=0. \label{e6} \ee
Here, $1/\lambda_d^2=1/\lambda_{dse}^2 + 1/\lambda_{dfe}^2 + 1/\lambda_{dfi}^2$. The second condition in (\ref{e5}) implies the absence of longitudinal modes in the flowing plasma alone.
In the ion acoustic frequency regime this dispersion equation yields \citep{v1, v2, vld2} the flow driven (current-less) instability with a drift-speed threshold far below
that of the classic electron-current driven kinetic instability. It will be shown here that even the ion plasma wave can become growing for rather realistic plasma parameters which could be used in actual plasma experiments. 
From Eq.~(\ref{e6}) the IP wave frequency is $\omega^2\simeq \omega_{psi}^2\left(1+ 3 k^2 \lambda_{dsi}^2\right)$.
The imaginary part of the frequency is obtained as before,   $\gamma\simeq -\mbox{Im} \Delta/(\partial \mbox{Re} \Delta/\partial \omega)_{\omega\equiv \omega_r}$, and clearly it can change  sign for large enough speed of the flowing plasma $v_{f0}$. The electron contribution to the term with the Doppler shifted frequency $\omega_0\equiv \omega- k v_{f0}$ in  the imaginary part of Eq.~(\ref{e6}) is clearly negligible; the ratio of the electron and ion terms associated with $\omega_0$ is  $(m_e/m_{fi})^{1/2} (T_{fi}/T_{fe})^{3/2}$
and it is typically much below unity.

We shall calculate the frequency and the growth rate from Eq.~(\ref{e6})  taking the static argon plasma parameters as in Sec.~\ref{sec2}.
The general frequency (which does include the IA wave range as well) is given by
\be
\omega^2\!=\!\frac{\omega_{psi}^2}{2\left[1+ 1/(k^2\lambda_d^2)\right]}\!\left\{\!1\!+\!\left[\!1\!+\! \frac{12 k^2 \vtsi^2}{\omega_{psi}^2}\!\left(\!1\!+\!\frac{1}{k^2\lambda_d^2}\!\right)\!\right]^{\!1/2}\!\right\},
\label{e8}
\ee
where
\[
 \frac{1}{k^2\lambda_d^2}\equiv \frac{1}{k^2\lambda_{dse}^2}
+\frac{1}{k^2\lambda_{dfe}^2} +\frac{1}{k^2\lambda_{dfi}^2}.
\]
In the absence of the $f$-components this  frequency reduces to  Eq.~(\ref{f}). Because of the term $1/(k^2\lambda_d^2)$, both static and flowing  components can now contribute to the real part of the frequency and can change its features, unless the contribution of $1/(k^2\lambda_d^2)$ is negligible. Hence, to remain in the IP frequency range, it is necessary to take wavelengths such  that $1/(k^2\lambda_d^2)<1$ or that it is of the order of unity.

The imaginary part of the frequency becomes
\[
\gamma=-\left(\frac{\pi}{8}\right)^{1/2} \frac{\omega_r^4}{\omega_{psi}^2 \left(1+ \frac{\displaystyle{6 k^2 \vtsi^2}}{\displaystyle{\omega_r^2}}\right)} \left[ \frac{\omega_{pse}^2}{k^3\vtse^3} \right.
\]
\[
\left. + \left(1- \frac{k v_{f0}}{\omega_r}\right) \left(\frac{\omega_{pfe}^2}{k^3\vtfe^3} + \frac{\omega_{pfi}^2}{k^3\vtfi^3}\right)\right.
\]
\be
\left.
+ \frac{\omega_{psi}^2}{k^3\vtsi^3}\exp\left(-\frac{\omega_r^2}{2 k^2 \vtsi^2}\right)\right].
\label{e9}
\ee
The damping (and growth rate) $\gamma$ is calculated in terms of the flowing plasma speed $v_{f0}$ by taking the same parameters for both the static  and flowing argon plasma as in Figs.~\ref{f1}, \ref{f2}, that is $T_{se}=T_{fe}=5\cdot 10^4$ K, $T_{si}=T_{fi}=0.1 \,T_{se}$,  and $n_{s0}=n_{f0}=10^{15}$ m$^{-3}$  (this would imply that one plasma is accelerated into another one externally, e.g.,  by an externally applied electric field). Such a choice of parameters is only for simplicity, i.e., to have the conditions used in the derivations satisfied in the simplest way, and to avoid solving the general dispersion function (\ref{pdf}) numerically. We  fix two  wavelengths $\lambda=0.001$ m, and $\lambda=0.03$ m, that correspond to the IP and IA wave regimes, respectively.

\begin{figure}
 \centering
\includegraphics[height=6cm,bb=19 16 275 218,clip=]{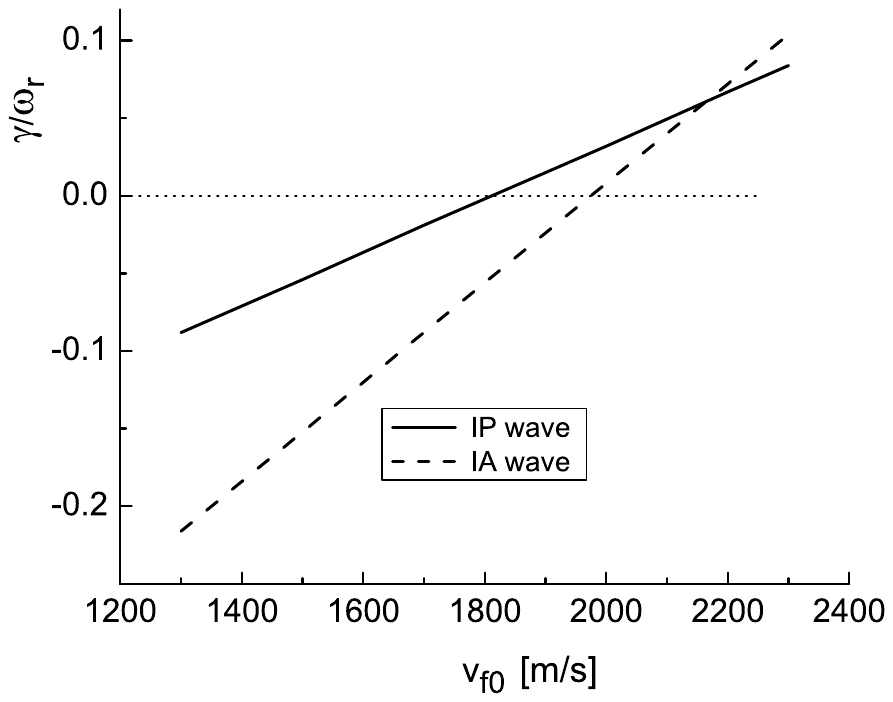}
\caption{\label{f3}Transition from damping to growing, for the IA wave ($k \lambda_d=0.03$), and the IP wave  ($k \lambda_d=0.88$), with increased speed of the flowing plasma.    }
\end{figure}

 The result is presented in Fig.~\ref{f3}. For the IP wave the frequency from Eq.~(\ref{e8}) is $\omega_r=1.16 \omega_{psi}\simeq 7.7\cdot 10^6$ Hz, while for the given IA wave $\omega_r=0.046 \omega_{psi}\simeq 3\cdot 10^5$ Hz. Observe that for the given parameters the acoustic speed in the static plasma
 $v_s\equiv c_s (1+ 3 T_{si}/T_{se})^{1/2}=3662$ m/s, where  $c_s^2= \kappa T_{se}/m_{si}$. From the graph it is seen that the IA wave becomes growing around  $v_{f0}>2000$ m/s, and the IP wave around  $v_{f0}>1800$ m/s. Hence,  in both cases this is far below the acoustic speed of the static plasma.  The drift-speed threshold of the IPW is lower because it has a smaller phase velocity than the IAW.

The presence of additional $f$-species changes both modes, but differently. The line for the IP wave in Fig.~\ref{f3} is above the IA line in the part  of the graph where the mode is damped (which is just the opposite to the case presented in Fig.~\ref{f2} where the IP wave damping is always stronger). So the IP wave  becomes destabilized easier in the presence of additional electron and ion components (in fact this is mainly due to additional ion species), although its growth rate for $v_{f0}>2200$ m is below the IA wave growth rate.

The conditions (\ref{e5}) used in derivations are not perfectly satisfied so that qualitative analysis might be replaced with more exact numerical solution
of the plasma dispersion function (\ref{pdf}), in particular in application to plasmas with hot ions like in the solar corona, but this is beyond the scope of this work. Note also that instead of  conditions  (\ref{e5}) we may have the following limits: $k \vtsi \ll |\omega|\ll k \vtse$, and $ k\vtfi\ll  |\omega - k v_{f0}|\ll k
\vtfe$. The dispersion equation obtained in such a manner can easily be derived; for the IA wave frequency range it is discussed in \citet{v1}.

\section{Summary}
The ion plasma wave is rarely studied  in the literature. Most likely this is the result of the fact that  in the simple electron-ion plasma, and with increased ion temperature,
the mode is (kinetically) more strongly damped than the ion acoustic wave. These features are presented in Figs.~\ref{f1},~\ref{f2}. On the other hand, the mode develops at spatial scales below the electron Debye radius, at which coherent and organized ion motion due to electrostatic force is usually not expected and the mode has thus remained out of the focus of researchers although it has been experimentally verified \citep{ba,ba2}.
 However, in case of a mixture of several ion species, the damping of the IP wave can be  reduced,
and in fact it can become lower than the damping of the IA wave in such an environment. This is shown quantitatively in Sec.~\ref{sec3} in the discussion related to Fig.~\ref{f3}.   Within kinetic theory, collisions additionally damp the IP mode but this damping can be smaller  than the damping of the IA mode for the same parameters.
In a fluid description, the ion plasma wave can be discussed without losing any physics by assuming the electrons as a completely fixed background.

In interpenetrating plasmas containing free energy in the flowing plasma component, the IP mode can become growing due to purely kinetic effects. Such an instability,  shown in Sec.~\ref{sec3}, develops for the speed of a flowing plasma that is well below the acoustic speed in the standing plasma. In addition, the IP mode is shown to be destabilized more easily that the IA mode.

The analysis presented in the work suggests that the ion plasma wave may be much more abundant than expected. In the solar wind environment at 1 AU distance from the Sun, this wave may develop  at wavelengths below $2\pi \lambda_{de}\sim 70$ m, and with frequencies of the order of a few kHz. For the magnetic field \citep{crav} of around $7$ nT  this yields $\lambda_{de}/\rho_e\simeq 0.01$ while in the same time  $\rho_i/\lambda_{de}\simeq 4000$. So for both ions and electrons  involved in the IP wave motion the magnetic field plays no role. In the solar corona, the IP mode implies wavelengths of a few centimeters or shorter, and frequencies in the range of $10^7$ Hz or higher, and it is again unaffected by the magnetic field, regardless of direction of propagation.

The short wavelength of the wave implies that it may propagate even in plasmas with the magnetic field, and at any angle with respect to the magnetic field vector. In such plasmas with  magnetic field, the electrons can affect the IP wave behavior on the condition that their gyro-radius is shorter than the Debye radius, or equivalently when the electron plasma frequency is below the electron gyrofrequency $\omega_{pe}<\Omega_e$. This is shown in Sec.~\ref{sec5}.

 In case of a growing mode, like the one excited in interpenetrating plasmas discussed in Sec.~III,  the energy of the wave can be channeled into internal (kinetic) plasma energy by the same stochastic heating mechanism known to work  in the case of the ion acoustic wave. This mechanism is described and  experimentally verified for the IA wave in \citet{sk1, sk2}. This means that the mentioned macroscopic flows in the solar atmosphere, as an example of interpenetrating plasmas that drive the IP wave unstable, may directly heat the solar plasma. In the case of IPW propagating along the magnetic flux tubes in the solar atmosphere, there may be refraction and self-focusing of the wave front in case of a greater density at the external regions of the tube. Such a ducting effect is known to play role in the electromagnetic wave propagation \citep{be, vld2} as well. For the IP wave this may result in greater amplitudes and such a linearly focused unshielded electric field of the wave may cause a more efficient  acceleration and energization of particles, similar to the case of experimentally amplified IP waves \cite{jon}.

\vfill

\pagebreak

\end{document}